\documentclass[showpacs,preprint,amsmath,amssymb]{revtex4}

\usepackage{graphicx}
\usepackage{dcolumn}
\usepackage{bm}

\begin{document}

\title{Photonic band gap and x-ray optics in warm dense matter}

\author{S. Ku$^{1}$\footnote[1]{Electronic address: {\sf sku@cims.nyu.edu}}\footnotemark[2],
S. Son$^{2}$\footnote[2]{These authors equally contributed to this work.}, and
Sung Joon Moon$^{3}$\footnote[3]{Current address: 28 Benjamin Rush Ln. Princeton, NJ 08540}}
\affiliation{$^{1}$Courant Institute of Mathematical Sciences, New York University, New York, NY 10012\\
$^{2}$18 Caleb Lane, Princeton, NJ 08540\\
$^{3}$Program in Applied and Computational Mathematics, Princeton University, Princeton, NJ 08544}
\date{\today}

\begin{abstract}
Photonic band gaps for the soft x-rays, formed in the periodic structures of
solids or dense plasmas, are theoretically investigated.
Optical manipulation mechanisms for the soft x-rays, which are based on these
band gaps, are computationally demonstrated.
The reflection and amplification of the soft x-rays, and the compression and
stretching of chirped soft x-ray pulses are discussed.
A scheme for lasing with atoms with two energy levels, utilizing the band
gap, is also studied.
\end{abstract}


\maketitle

\section{Introduction}

The electromagnetic (E\&M) waves impinging into a photonic band gap (PBG) material
(also known as the photonic crystal), consisting of periodic dielectric or
metallo-dielectric structure, exhibit similar behavior as electron waves in
semiconductor crystals~\cite{nature,Yablo,John}.
The existence of a stop (or forbidden) band forbidding the propagation of the E\&M waves
of certain wavelength is a direct analogy of the semiconductor band gap for the electron
waves, which leads to a number of phenomena that can be used to manipulate the E\&M waves.
Over the last two decades, a great progress has been made in developing the techniques
to manipulate the visible light waves using the PBG materials~\cite{PCbook}.

A few ways controlling and manipulating the soft x-rays have been proposed~\cite{Xbook},
including the Bragg diffraction occurring in metallic multi-layers~\cite{xrays,xrays2}. 
The photonic band gap materials interacting more strongly with the x-rays than the currently
available metallic multi-layers might be available soon,
based on the recent advances in the compression technology, as well as the development of
powerful energy sources such as free electron lasers~\cite{Free,Free2}.
It is becoming more meaningful to consider possible applications of such a strong PBG material
to optically manipulate the x-rays. 
A recent work suggests that powerful x-ray pulses may be stretched and compressed
using a relic lattice of dense plasmas~\cite{Fisch}.

In this paper, we theoretically study the PBG materials that might be created in dense plasmas
in near future or in metallic multi-layers that are already available in the laboratory,
and propose a few ways to manipulate an intense soft x-ray pulse, utilizing the PBG properties.
We consider soft x-rays of short durations, of the order of a few to tens of femtoseconds,
impinging into a dense plasma or metallic layers of alternating densities.
We discuss the mechanisms enabling the reflection of an x-ray pulse and the stretching
or compression of a chirped pulse, making use of the strong convexity of the dispersion
relation around the stop band (Secs. II and III).
The stretching and compression requires a relatively long pulse duration, yet the maximum
duration is limited by the inverse bremsstrahlung rate. 
We show that the population inversion is possible only with two energy levels.
We demonstrate that the laser can be amplified by controlling the spontaneous and
the stimulated emission around the band edge (Sec. IV), and conclude the paper (Sec. V).

\section{Photonic band gap}

We consider a plasma of a high conduction electron density which varies periodically in one direction ($x$-direction).
We denote the alternating thicknesses of the plasma by $d_A$ and $d_B$, which are
a few to 100 nm, and the conduction electron density in each region by $n_{A}$ and $n_{B}$
($n_B > n_A$), which are in the range of $10^{23}/\mathrm{cc}$ to $10^{24}/\mathrm{cc}$.  
The wavelength of the x-ray to be considered is comparable to $2(d_A+d_B)$.
For convenience, we introduce the notations
$\delta \omega_{\mathrm{pe}}^2 = 4\pi(n_B-n_A)e^2/m_e$, $\omega_B^2 = 4\pi n_B e^2/m_e$,
and $\omega_A^2 = 4\pi n_A e^2/m_e$, where $e$ and $m_e$ are the electron charge and mass, respectively.

As to be shown below, a higher value of the density difference, $\Delta n \equiv |n_B-n_A|$,
is preferred, as it exhibits a stronger band gap effect. 
For example, the conduction electron density of AL is $1.8 \times 10^{23} / \mathrm{cc} $,
and $\Delta n \cong 10^{23} / \mathrm{cc}$ is demonstrated to be achievable by the vacuum
deposition of the metallic multi-layers~\cite{xrays, xrays2}.
Even though a much higher electron density (up to $n_A (n_B) \cong 10^{26}/ \mathrm{cc} $)
might be obtained through the contemporary compression technologies~\cite{lindl},
it is technically very difficult, if not impossible, to create a periodic structure
of $\Delta n \gg 10^{23} / \mathrm{cc}$ that may last a long enough time for any practical use.
 
One highly plausible way to achieve a structure of a high $\Delta n$ is as follows.
One may create alternating layers of two metals and then quickly ionize it
(in the order of tens of femtoseconds) using an intense x-ray. 
A recent experiment shows that such a rapid ($<$ 15 femtoseconds) photo-ionization
creates a ``crystalline'' plasma~\cite{exp1}.
The conduction electron density in each metal increases as the ionization progresses,
and a considerably high $\Delta n$ ($\cong 10^{23} / \mathrm{cc} $) can be achieved,
depending on the atom density, the atomic level, the ionization cross-section of each
metal, and the power and the wavelength of the ionizing x-ray.
It might be even possible to reach $\Delta n > 5 \times  10^{23} / \mathrm{cc}$
if some of the inner-shell electrons in the heavy metal ion get ionized. 
Once the above condition is met, the disintegration time ($>\sim 10^{-11} \sec$)
exceeds the timescale a light pulse interacts with the PBG materials
($<\sim 10^{-14} \sec$).  

Another possible way would be compressing or ionizing two metal layers
at different rates by a shock wave.
However, as the compression time is likely to be longer than a picosecond,
the Rayleigh-Taylor instability may develop before the periodic structure
can be used by anything else.
One more method  is to utilize the ponderomotive potential of two-beating x-rays~\cite{PPC}.
In this case, the Raman scattering from the x-rays might limit the number of
layers that could be created. 
As discussed, there are many experimental challenges to be overcome before such a periodic
structure can be created and used for practical applications.
Assuming such issues would be overcome by the further developments of
the experiment, we concentrate on the theoretical aspects in the following.

We begin by considering the Maxwell's equation:
\begin{equation} 
\frac{1}{c^2}\frac{\partial^2 A_y}{\partial t^2 } = -\frac{4 \pi n_e(x) e^2}{c^2 m_e}A_y + \frac{\partial^2 A_y}{\partial x^2},
\label{eq:maxwell}
\end{equation} 
where $E_y = (-1/c) (\partial A_y/\partial t)$,
$B_z = \partial A_y/\partial x$, and $j_y = n_e(x)e^2 A_y /m_e c$.
The spatial variation of the density in the $x$-direction is emphasized by $n_e(x)$.
The E\&M wave propagates in the $x$-direction with the electric field (current)
in the $y$-direction and the magnetic field in the $z$-direction.
The response of the plasma would depend on $n_e(x)$, from which the band
gap structure arises.

\begin{figure}
\scalebox{0.33}{\includegraphics{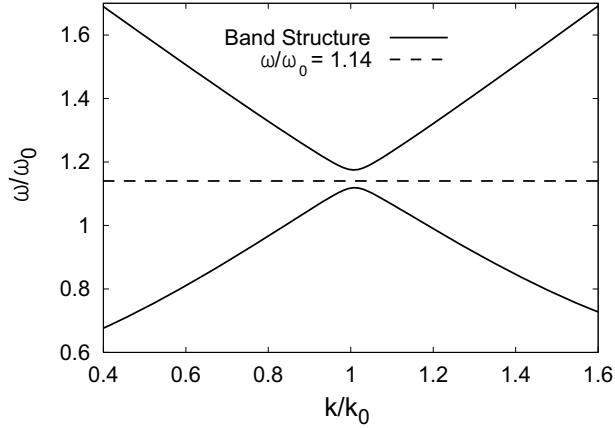}}
\caption{\label{fig:band}
The photonic band structure formed in a dense plasma, where $k_0 = 2\pi/(2d_A + 2d_B) = 2\pi/(20 \mathrm{nm})$ and $\omega_0 =c k_0$ ($d_A=d_B=5 \ \mathrm{nm}$, $ n_A = 5.58 \times 10^{23}/\mathrm{cc}$, $n_B=2n_A$).
Dashed line, $\omega / \omega_0=1.14$, is drawn for the guidance of the eyes.}
\end{figure}

\begin{figure}
\scalebox{0.65}{\includegraphics{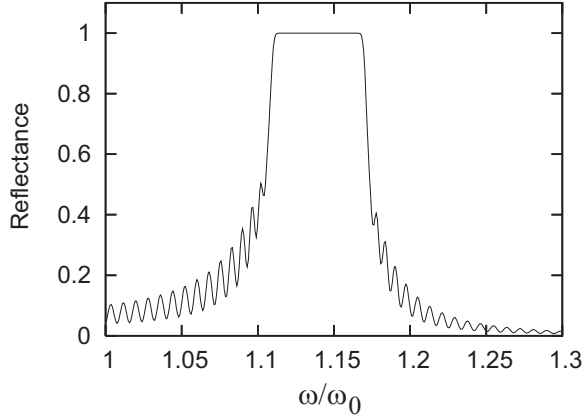}}
\caption{\label{fig:reflection}
Numerically obtained reflectance as a function of the mid-wave frequency $\omega$ of a 6.6 femtosecond Gaussian pulse, where $\omega_0 = 9.42 \times 10^{-16}/\sec$.
A PBG material extended over 100 nm is simulated.}
\end{figure}

When a single-frequency x-ray pulse impinges into the material ($x$-direction),
the propagation of the pulse is forbidden if the frequency lies within the
photonic band gap. 
For example, the dispersion relation of the band gap material with
$d_A=d_B=5~\mathrm{nm}$, $n_A = 5.58 \times 10^{23}/\mathrm{cc}$,
and $n_B=2n_A$ exhibits a band gap at $\omega=1.14\omega_0$ (Fig.~\ref{fig:band}),
where $k_0 = 2 \pi /\lambda_0 = 2 \pi /(2d_A +2d_B) =2 \pi /(20~\mathrm{nm})$ and $\omega_0 = c k_0= 9.42 \times 10^{16} /\sec$.
From here on, the wave vectors are scaled by $k_0$.
The band structure acts as an x-ray mirror for a range of frequencies;
numerical integration of Eq.~(\ref{eq:maxwell}), using the pseudo-spectral scheme
with the Crank-Nicolson time-stepping method, shows that pulses of the frequencies
in the stop band get reflected from the PBG material.
For instance, consider a 6.6 femtosecond Gaussian pulse propagating in the $x$-direction, given the same physical parameters used in Fig.~\ref{fig:band}.
The pulse can be completely specified by the mid-wave vector $k_m$ in the Gaussian pulse.
We consider a PBG material extended over 100 nm, which corresponds to $10(d_A+d_B)$.
The pulse impinging into the PBG material gets transmitted or reflected, depending on
the wave vector.
We observe that the total reflection occurs when $k_m$ lies between $\sim 1.1$
and 1.2 (Fig.~\ref{fig:reflection}).
A pulse partially overlapped with the gap gets partially transmitted and reflected.
We observe that the transmitted pulse travels much slower than the one reflected.

Using a simple perturbation analysis, one can see that the band size is
$\delta \omega \sim \delta \omega_{\mathrm{pe}}^2/\omega_0$.
The minimum Gaussian pulse duration for the total reflection to occur is estimated
to be $1/( \delta \omega)$.
When $\delta \omega_{\mathrm{pe}} \approx  \omega_{\mathrm{pe}}$, ten times shorter
pulse can get reflected by the material of ten times higher electron density.
The gap size is proportional to $\Delta n$.
The stronger the interaction of the x-ray with the PBG material is, the higher $\Delta n$ is. 
The skin depth for the attenuation in the band gap is estimated as
\begin{equation}  
L_s = \frac{c}{\delta \omega} = c \frac{\omega_0}{\delta \omega_{\mathrm{pe}}^2} \mathrm{.} \label{eq:skin}
\end{equation}  
The minimum number of layers for the reflection to occur is
$(\omega_0/ \omega_{\mathrm{pe}})^2$.

\section{Compression and Stretching}
\begin{figure}
\scalebox{0.5}{\includegraphics{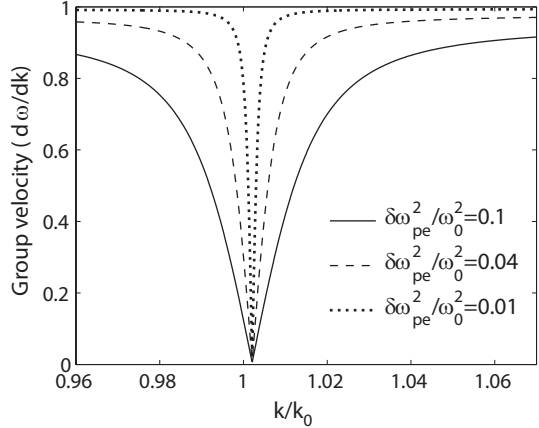}}
\caption{\label{fig:10}
Numerically obtained group velocity as a function of the wave vector $k$
when $\omega_B^2=2\omega_A^2$, for the three cases where
$\delta \omega_{\mathrm{pe}}^2/\omega_0^2$ is 0.1 (solid line), 0.04 (dashed line),
and 0.01 (dotted line).
The group velocity in the $y$-axis is scaled by the velocity of light.
}
\end{figure}

While a pulse of the forbidden frequency gets reflected, a pulse of the frequency
slightly above or below the band gap would be transmitted.
As $d\omega/dk \sim 0$ around the band gap and $d^2\omega /dk^2 \gg 0 $
($d^2\omega /dk^2 \ll 0$) slightly above (below) the gap, the higher frequency component
of the pulse, slightly below the band gap, travels more slowly than the lower frequency
component.
Consequently, the pulse would be stretched or ``positively frequency-chirped''.
The opposite would occur for a pulse of the frequency above the gap.
We compute the group velocity $d\omega/dk$ in a range of $k$ 
for three different values of $\delta\omega_{\mathrm{pe}}^2/\omega_0^2$,
when $\omega_B^2=2 \omega_A^2$ (Fig.~\ref{fig:10}).
There is a small window of the wave vector $ \delta k = 0.5
(\delta\omega_{\mathrm{pe}}/\omega_0)^2 k_0$ where the group velocity
deviates from the velocity of the light $c$ in a homogeneous plasma. 
In order to have a wave packet with a well-defined group velocity that is
considerably different from $c$, the band width of the pulse should be much smaller
than the above band width, which is estimated to be
$\delta k = 0.1 (\delta \omega_{\mathrm{pe}}/\omega_0)^2 k_0$.  
The minimum pulse duration is given by 
\begin{equation}
\tau_m = 10 \left( \frac{\omega_0}{\delta \omega_{\mathrm{pe}}} \right)^2 \frac{1}{\omega} \label{eq:minimum} \mathrm{,}
\end{equation}
exhibiting an inverse relationship between the coherent pulse and $\delta \omega_{\mathrm{pe}}$.
For the stretching and compression, the pulse length could be twice of what is given
in Eq.~(\ref{eq:minimum}).
Denoting the thickness of the PBG material as $D$, the achieved stretching would be given
as $\left( D/v_{\mathrm{min}} - D/v_{\mathrm{max}} \right)/c$, where $v_\mathrm{max}$
($v_\mathrm{min}$) is the maximum (minimum) velocity in the components of the pulse.
The compression also would work the same way.

The pulse length for the stretching or the compression is limited by the size of the region
satisfying
$d\omega/dk \sim 0$.
A better stretching or compression can be achieved with a larger value
of $\delta \omega_{\mathrm{pe}}/\omega$, which is bounded by $\omega_{B}/\omega$.
In the estimation here and the following, we assume that $\omega \simeq \omega_0$,
as we are interested only in the wave packets near the band structure.
While a longer pulse is preferred for the better band gap effects, such a pulse would
be dissipated further via inverse bremsstrahlung.
The inverse bremsstrahlung in the warm dense matter is given as
$\nu(\omega) = (2 Z_i  e^4 m_e / 9 \pi \hbar^3) (\omega_{pe}^2/\omega^2) Q^{-3/2}$,
assuming $ Q =  (\hbar \omega / 2 m_e v_F^2) > 1 $ and
$2e^2E_y^2/m\hbar \omega^3 < 1$, where $Z_i$ is the ion charge number
and $v_F$ is the Fermi energy~\cite{shima}.
We take a half of the above formula assuming most of the inverse bremsstrahlung
is accounted for by the denser layer (the region $B$ in our notation).
Then the above equation is estimated to be
\begin{equation}
 \nu_B \equiv \frac{1}{2} \nu = 1.46 \times 10^{15} Z_i q^{-3/2} \frac{\omega_{\mathrm{pe}}^2}{\omega^2} ~~~\mathrm{sec}^{-1} . 
\end{equation}
The product of the minimum duration of the pulse and the inverse bremsstrahlung decay rate is given as $\tau_m \nu_B = 1.46 \times 10^{16} Z_i q^{-3/2} / \omega$, assuming $\delta \omega_{\mathrm{pe}} \simeq \omega_{\mathrm{pe}}$.
It should be much smaller than the unity, in order to facilitate enough time for the compression or the stretching.  
For example, when $n_e = 10^{24}/ \mathrm{cc}$, $Z_i=5$, $\delta \omega_{\mathrm{pe}}\approx \omega_{\mathrm{pe}}$, and $\omega_{\mathrm{pe}}/\omega = 0.1$, the inverse bremsstrahlung time goes up as large as 50 femtoseconds and $\tau_m \nu_B$ = 0.03.
We might be able to, in principle, stretch a pulse to the inverse of the bremsstrahlung rate.

The above-mentioned compression and stretching mechanisms may be combined to facilitate such applications as the ``Chirped Pulse Amplification''~\cite{french} of the x-rays, given that a soft x-ray laser of the gain intensity of $\sim$ $60/\mathrm{cm}$ is already available~\cite{gain}.
Currently available soft x-ray amplification technologies typically use a laser in the visible light regime in order to achieve the population inversion from the collisional excitation~\cite{laser,laser1,gain}, and the maximum gain is obtained when the duration is of the order of picosecond.
On the other hand, a much shorter pulse is required for many ultra-short and high power
laser experiments and applications.
Our stretching and compression mechanisms may meet such needs.
However, as shown in Eq.~(\ref{eq:minimum}), the minimum pulse duration is much
larger than the skin depth given in Eq.~(\ref{eq:skin}), which requires the physical
dimension of the photonic band gap material to be at least 100 times bigger than
the skin depth for a sufficient compression or stretching to occur.
Such a physical dimension might be too big for the dense plasmas, due to the
ionization power limit and/or the uniformity requirement.  
Such a relatively large structure of the PBG material can be easily achieved in metallic multi-layers. 
Considering such practical limitations,
the compression and stretching proposed in this section would be
more relevant to the low-energy soft x-ray pulses in the metallic multi-layers,
since the pulse of a low frequency interacts with the metal more strongly.

\section{Lasing with two-level atoms}

In the currently available x-ray amplification schemes~\cite{laser,laser1,gain}, the gain intensity is limited by the carrier density (the ion density with the population inversion) of the amplifying media.
Consequently, a higher ion density is needed for a better amplifying power; however, in order to achieve a population inversion in such schemes, the ion density (hence the free electron density) should remain under some critical density above which the pumping laser field no longer penetrates the plasma.
This constraint on the ion density limits the gain intensity to be maximized.

In contrast to a common, intuitive belief that the spontaneous emission rate is an intrinsic property of a material, and that the photons could be manipulated only after the emission, it was shown both theoretically and experimentally that the rate of the spontaneous emission and the stimulated emission can be controlled by varying the background photon state density~\cite{Yablo,Purcell,Seleny}.
Along this line, we consider an atom with two electron energy levels $p$ and $q$.
Via the spontaneous emission, the stimulated emission or absorption, an electron in the atom may jump between the states $p$ and $q$ by absorbing or emitting a photon of the energy $\hbar \omega_{pq} = E_p - E_q$. 
The transition rate, from the Fermi's Golden Rule, is
\begin{equation}
\gamma \propto (2\pi/\hbar) \langle p|\mathbf{E}|q\rangle^2 \rho(\omega_{pq}) \mathrm{,}
\end{equation}
where $\rho$ is the density of the available photon states.
As was shown experimentally, the spontaneous emission is prohibited when an atom is located inside the cavity where photon states are unavailable~\cite{Purcell}.
Similarly, if an atom is located inside the PBG material and there are no available photon states for $\hbar\omega$, the spontaneous or the stimulated emission would be forbidden since $\rho(\omega)=0$.
On the other hand, if $\omega_{pq}$ is just outside the band gap, there would be more available states for a photon, as the photon density is proportional to $dk/d\omega$, which is very large.  

Consider a PBG material with a stop band $\omega_0 - \delta \omega < \omega_{pq} <  \omega_0 + \delta \omega$, where the region $A$ contains many atoms with the two-level energy for electrons, as described above.
When a powerful x-ray source is pumped in the $y$-direction into the PBG material, an electron in the atom at the state $p$ or $q$ would absorb or emit a photon via the spontaneous or the stimulated emission, and absorb a photon via the resonant absorption.
Denoting the average occupation numbers per atom by $n_p$ and $n_q$, the time evolution is described by
\begin{eqnarray}
\frac{d n_p}{dt} &=& -\alpha n_p + \beta n_q - \kappa n_p \nonumber -\Gamma_p \\
\frac{d n_q}{dt} &=& \alpha n_p -\beta n_q +\kappa n_q - \Gamma_q \nonumber
\label{eq:2}\mathrm{,}
\end{eqnarray}
where $\alpha$ is the stimulated emission rate coefficient, $\beta$ is the resonant absorption rate coefficient, and $\kappa$ is the spontaneous emission rate coefficient, $\Gamma_p$ ($\Gamma_q$) accounts for other decay processes such as the Auger process.
The spontaneous emission rate is usually larger than a picosecond.
We assume that $\alpha \gg \kappa, \Gamma_p $, and  $\beta \gg \kappa, \Gamma_q$,
which is the case if many enough x-ray photons are pumped during a very short time so that the average
rate of the transition is much higher than the spontaneous emission rate and
the decay due to other processes.
Denoting the pumping pulse duration by $\tau_{\mathrm{pump}}$, the number of photon
in the pump by $N_{\mathrm{pump}}$, and the volume of the plasma to be pumped by $V_{\mathrm{pump}}$,
the transition rates $\alpha n_p $ and $\beta n_q$, assuming $n_p,n_q\cong 0.5$,
are roughly estimate as $ N_{\mathrm{pump}} \sigma c / V_{\mathrm{pump}}$,
where $\sigma$ is the ionization cross-section. 
Then the condition of the fast stimulated emission (absorption) compared to the decay
turns out to be
\begin{eqnarray} 
 V_{\mathrm{pump}} &\ll& N_{\mathrm{pump}} \sigma c / \kappa, \ \ \mathrm{and} \nonumber \\ \nonumber \\
 V_{\mathrm{pump}} &\ll& N_{\mathrm{pump}} \sigma c / \Gamma_{p,q}   \mathrm{.}     \\  \nonumber
\end{eqnarray}
We assume the ionization cross-section to be $\sigma = 10^{18}\  \mathrm{cm}^2$
in the following discussion:
The number of photons per pulse of the Linac coherent light source in Stanford is
$10^{12}$ in a few tens or hundreds of femtoseconds.
Then the maximum volume $V_{\mathrm{pump}}$ is estimated to be $3 \times 10^{-6} \mathrm{cm}^3 $, assuming the decay rate is $10^{12} / \sec$. 
The decay due to the Auger process would be very fast ($10^{14} / \sec$),
yet it could be greatly suppressed by the degeneracy of the plasma.
Assuming a decay rate of $10^{14}/ \sec $, the maximum volume $V_{\mathrm{pump}}$ is
estimated to be $3 \times 10^{-8} \mathrm{cm}^3 $. 
This is achievable, considering the fact that the focusing could be as small as
$10 \ \mu \mathrm{m}$. 
Also, in a strong free electron laser where the number of photons per pulse is
between $10^{13}/ \sec $ and $ 10^{14} / \sec$, the volume can be considerably bigger.   
For such a very fast decay rate as $10^{15}/ \sec$, the pump pulse duration should be
also of order of femtosecond, in order for the pump condition to be achieved.
Assuming the number of photons per pulse is $10^{14}$, the maximum volume
$V_{\mathrm{pump}}$ is estimated to be $3 \times 10^{-7} / \sec$.

Let us assume that, by focusing the photon beam, the stimulated emission and the absorption rates are considerably higher than the other decay rates. 
Then a steady state between the densities $n_p$ and $n_q$ would be arrived so rapidly that $ \alpha n_p + \beta n_q = 0 $.
The relationship $\alpha = \beta $ would hold in a vacuum, due to the Fermi's
Golden Rule, and the population inversion cannot occur.
However, in the PBG material, the inequality $\alpha < \beta $ can be satisfied, due to the inhibition of the stimulated emission.
A photon of the energy $\hbar \omega_{pq}$ can be emitted in the most directions,
yet it cannot be emitted in the $x$-direction.

Now we consider launching a seed x-ray pulse in the $x$-direction, of the frequency slightly outside the stop band.
There are two benefits in this approach.
First, the density of the states available for the photons is high and $dk/d\omega \gg 1/c$, so that the stimulated emission can be asymmetrically enhanced compared to the resonant absorption.
This asymmetry would enable the lasing action even without the population inversion, as long as the occupation density per atom reaches its steady state due to the pumping.
Second, the group velocity of the pulse is very small, making the pulse spend more time in the PBG to obtain a higher gain, as was shown in Ref.~\cite{edgelaser}.
The lasing gain intensity is given by 
\begin{equation} 
 g = \left( (1+\epsilon)  n_p - n_q \right) c \sigma \mathrm{,}
\end{equation}
where $\epsilon $ is the enhancement factor for the stimulated emission due to
the denser photon state and $\sigma$ is the absorption cross-section.
The population inversion and a positive value of $\epsilon$ could considerably enhance the lasing gain in dense plasmas.
One constraining factor is that the gaining intensity should be bigger than the inverse bremsstrahlung
\begin{equation}
g/\nu = \frac{2.11}{Z_i} \frac{n_{\mathrm{eff}}\sigma}{Q^{-3/2}}
  \frac{\omega^2}{\omega_{\mathrm{pe}}^2}\mathrm{,}
\end{equation}
which is larger than the unity, where $n_{\mathrm{eff}} =(1+\epsilon) n_p - n_q$ is scaled by $10^{-23}/\mathrm{cc}$ and $\sigma$ by $10^{-18}\mathrm{cm}^2$.
If $n_{\mathrm{eff}} = 10^{22}/\mathrm{cc}$ is achieved, $g/\nu$ becomes larger than the unity for $\omega/\omega_{\mathrm{pe}} > 5 $, even for a very large $Z_i$.
Then the gain would be as large as $10000/\mathrm{cm}$ for the absorption cross-section of $10^{-18}\  \mathrm{cm}^2$.
This gain intensity is orders of magnitude higher than the currently available conventional x-ray laser amplification technologies.

The minimum size of the photonic band gap needed for the amplification is the skin depth estimated in Eq.~(\ref{eq:skin}), where a photon experiences the presence of the photonic band gap. 
This size is much smaller than what is needed in the compression or the stretching,
so that the amplification can be operated either in metals or dense plasmas.

\section{Conclusion}

It is shown that the PBG formed in metals or the warm dense matters enables
manipulating a soft x-ray pulse of the wavelength in the range of 1 to 100 nm,
including the reflection, the compression, and the stretching.
For the low-energy soft x-ray pulses, the compression and the stretching would be
more feasible in metals than in dense plasmas, as the required physical dimension
of the PBG material is more realistic in the former.

It is also shown that the population inversion and a powerful laser amplification would
be possible if a prudent usage for the forbidden band gap is made to enhance or
inhibit the stimulated emission.
The amplification scheme proposed here can be equivalently operated either in metals
or dense plasmas.

While the proposal made here can be operated in the metallic multi-layers
that are already available in the laboratory, the size of the resulting
band gap is relatively small,
and a stronger photonic band gap material formed in dense plasmas is desired.
A considerable density variation of the conduction electrons between the layers
in the dense plasmas is needed to meet this goal, as
the band gap width is roughly proportional to the density difference between
the layers.
The creation of a periodic plasma structure of such a high density differences
would be the major hurdle before the applications discussed here are realized
in the laboratory experiment.
When it is achieved, the door to more applications of the PBG in dense plasmas
would open.
It would be interesting to see if the soliton compression~\cite{soliton},
the second harmonic generation~\cite{second}, and the surface
plasmon~\cite{ritchie} are relevant to dense plasmas. 


\end{document}